\documentclass[aps,prc,twocolumn,groupedaddress,showpacs,eqsecnum]{revtex4}
\usepackage{graphicx}
\usepackage{dcolumn}
\usepackage{bm}

\newcommand{\ran}{($\alpha$,n)}

\newcommand{\rng}{(n,$\gamma$)}

\newcommand{\rag}{($\alpha$,$\gamma$)}

\newcommand{\rpg}{(p,$\gamma$)}

\newcommand{\taN}{$^{180}$Ta}

\newcommand{\luN}{$^{176}$Lu}

\newcommand{\spro}{s-process}
\newcommand{\rpro}{r-process}
\newcommand{\ppro}{p-process}

\newcommand{\g}{$\gamma$}
\newcommand{\loK}{low-$K$}
\newcommand{\hiK}{high-$K$}
\newcommand{\Kmix}{{\it{K}}-mixing}

\begin{document}

\title{

Modification of nuclear transitions in stellar plasma by electronic processes:
$K$-isomers in $^{176}$Lu and $^{180}$Ta under s-process conditions
}

\author{G.\ Gosselin
}
\email[E-mail: ]{gilbert.gosselin@cea.fr}
\affiliation{
CEA, DAM, DIF F-91297 Arpajon, France
}

\author{P.\ Morel
}
\email[E-mail: ]{pascal.morel@cea.fr}
\affiliation{
CEA, DAM, DIF F-91297 Arpajon, France
}

\author{P.\ Mohr
}
\email[E-mail: ]{WidmaierMohr@t-online.de}
\affiliation{
Diakonie-Klinikum Schw\"abisch Hall, D-74523 Schw\"abisch Hall,
Germany
}
\affiliation{
Institute of Nuclear Research (ATOMKI), H-4001 Debrecen, Hungary
}

\date{\today}

\begin{abstract}
The influence of the stellar plasma on the production and destruction of
$K$-isomers is studied for the examples $^{176}$Lu and $^{180}$Ta. Individual
electromagnetic
transitions are enhanced predominantly by nuclear excitation by electron
capture, whereas the other mechanisms of electron scattering and nuclear
excitation by electron transition give only minor contributions. It is found
that individual transitions can be enhanced significantly for low transition
energies below 100\,keV. Transitions with higher energies above 200\,keV are
practically not affected. Although one low-energy transition in $^{180}$Ta is
enhanced by up to a factor of 10, the stellar transition rates from low-$K$ to
high-$K$ states via so-called intermediate states in $^{176}$Lu and
$^{180}$Ta do not change significantly under s-process conditions. The
s-process nucleosynthesis of $^{176}$Lu and $^{180}$Ta remains essentially
unchanged. 
\end{abstract}


\maketitle

\section{Introduction}
\label{sec:intro}
In general, nuclear properties like decay half-lives and radiation widths do
not depend on the electronic environment of the atomic nucleus. However, there
are several well-known exceptions. Obviously, electron capture decays are
affected by the number of available electrons, in particular $K$-electrons,
and thus the $K$-capture half-life depends on the ionization of the atom. A
second example are low-energy \g -transitions where the decay widths are
enhanced by additional conversion electrons. The present study focuses on
further effects that may affect nuclear transitions in a hot and dense plasma
that is found in the interior of stars: inelastic and superelastic electron
scattering and nuclear excitation by electron capture (NEEC)
\cite{Doo78,Gos07}; NEEC is the inverse process of the above mentioned
internal conversion (IC). Furthermore, even nuclear excitation by electron
transition (NEET) \cite{Mor04} may be important if matching conditions can be
achieved. 

As will be shown in this study, \g -transitions with relatively low energies
far below 1\,MeV are most affected by the surrounding hot and dense
plasma. Typical \g -transition energies for \rng , \rpg , and \rag\ capture
reactions are of the order of 1\,MeV and higher and are thus not significantly
affected by the stellar plasma. However, low-energy \g -transitions play an
important role in the production and destruction of low-lying isomers in the
astrophysical \spro . There are two astrophysically relevant examples for
heavy odd-odd nuclei where low-lying isomers exist because of the huge
difference of the $K$-quantum number between the ground state and the isomer:
\luN\ and \taN . The astrophysical transition rates between the \loK\ and
\hiK\ states in \luN\ and \taN\ may be affected by the temperature dependence
of the individual transitions.

The interesting astrophysical properties of \luN\ and \taN\ will not be
repeated here. The \spro\ nucleosynthesis of \luN\ and $^{176}$Hf and the
interpretation of the $^{176}$Hf/\luN\ ratio as \spro\ thermometer are
discussed in several recent papers (see \cite{Heil08,Mohr09,Gin09}, and
references therein). The open question on the nucleosynthetic origin of
\taN\ in various processes (\spro , \rpro , \ppro\ or $\gamma$-process,
$\nu$-process) and the survival probability of the $9^-$ isomer under the
corresponding conditions was also studied recently (see \cite{Mohr07} and
references therein).

The main subject of the present study is the temperature dependence of
individual transitions from an initial state $i$ to a final state $f$. This
general
temperature dependence should not be mixed up with the temperature dependence
of the stellar transition rates between \loK\ states and \hiK\ states in
\luN\ and \taN\ that are defined by low-lying so-called intermediate states and
their decay properties -- i.e.\ all possible transitions from these
intermediate states. It is obvious that changes in the individual transitions
-- as studied in this work -- do also affect the stellar transition rates.

The paper is organized as follows. In Sect.~\ref{sec:rate} some introductory
remarks on the nuclear structure of isomers are given, and the stellar
reaction rate between \loK\ states and \hiK\ states is defined. In
Sect.~\ref{sec:mod} the temperature dependence of individual transitions is
discussed. Results for selected individual transitions in \luN\ and \taN\ are
presented in Sect.~\ref{sec:res}, and their influence on the stellar
transition rate is discussed. Finally, conclusions are drawn in
Sect.~\ref{sec:summ}. As usual, we will give the ``temperature'' in units of
keV, i.e.\ the temperature $T$ is multiplied by the Boltzmann constant $k$
leading to the thermal energy $kT$.

\section{Stellar reaction rates}
\label{sec:rate}
\subsection{Nuclear structure}
\label{sec:struc}
The approximate conservation of the $K$-quantum number leads to a strong
suppression of direct transitions between so-called \loK\ and \hiK\ states in
heavy nuclei. As a consequence, the \loK\ $J^\pi = 1^-;K = 0$ state in
\luN\ at $E_x = 123$\,keV practically cannot decay to the \hiK\ $7^-;7$ ground
state. Instead, the \loK\ $1^-;0$ state forms an isomer that $\beta$-decays
with a half-life of $t_{1/2} = 3.66$\,h to $^{176}$Hf. The $\beta$-decay of
the $7^-;7$ ground state is also highly suppressed and has a long half-life of
about 38 giga-years, i.e.\ it is practically stable for the timescale of the
astrophysical \spro .  In \taN\ the roles of the ground state and the isomer
are exchanged: the \loK\ $1^+;1$ state is the ground state and has a short
$\beta$-decay half-life of about $8.15$\,h whereas the \hiK\ $9^-;9$ isomer at
$E_x = 77$\,keV is quasi-stable with $t_{1/2} > 7.1 \times 10^{15}$\,yr
\cite{Hul06}. Excitation energies, spins and parities, half-lives, and decay
properties are in most cases taken from the online data base ENSDF
\cite{ENSDF} that is based on \cite{Bas06} for \luN\ and \cite{Wu03} for \taN
; other data sources are stated explicitly.

Because of the strong suppression of direct transitions between the \loK\ and
the \hiK\ states, two species (a \loK\ one and a \hiK\ one) of such nuclei
like \luN\ and \taN\ have to be considered in nucleosynthesis calculations
(see e.g.\ \cite{Heil08}). Within each species, thermal equilibrium is
obtained on timescales of far below one second (e.g.\ explicitly shown in
\cite{Gin09} for \luN ). However, indirect transitions 
between the \loK\ and the \hiK\ states are possible via
so-called intermediate states (IMS) that are located at higher excitation
energies and have intermediate $K$-quantum numbers. Such IMS have been
detected experimentally by high-resolution \g -ray spectroscopy for
\luN\ \cite{Klay91a,Klay91b,Les91,Pet92,Dra10}, and an indirect proof for the
existence of IMS was obtained from various photoactivation studies
\cite{Ver70,Wat81,Nor85,Carr89,Carr91,Lak91,Lak95a,Lak95b,Van00,Kn05}. A
review of the results for \luN\ is given in \cite{Mohr09}. For \taN\ only
indirect evidence for the existence of IMS was derived from photoactivation
\cite{Bel99,Bel02,Lak00,Car89,Col88,Col90,Nem92,Nor84,Bik99,Sch94,Sch98,Loe96,Sch01,Loe03}. A
direct detection of IMS by \g -spectroscopy was not possible up to now, see
e.g. \cite{Dra98,Sai99,Dra00,Wen01}.

\subsection{Definition of astrophysical reaction rates}
\label{sec:def}
The stellar transition rate $\lambda^\ast$ for transitions from the
\loK\ to the \hiK\ species of heavy nuclei is approximately given by
\begin{eqnarray}
\lambda^\ast(T) & = & 
\int c \, n_\gamma(E,T) \, \sigma(E) \, dE \nonumber \\
& \approx &
c \sum_i n_\gamma(E_{IMS,i},T) \, I^\ast_\sigma(E_{IMS,i})
\label{eq:lam}
\end{eqnarray}
with the thermal photon density 
\begin{equation}
n_\gamma(E,T) = 
  \left( \frac{1}{\pi} \right)^2 \,
  \left( \frac{1}{\hbar c} \right)^3 \,
  \frac{E^2}{\exp{(E/kT)} - 1}
\label{eq:planck}
\end{equation}
and the energy-integrated cross section $I^\ast_\sigma$ under stellar
conditions for an IMS at excitation energy $E_{IMS}$
\begin{eqnarray}
I^\ast_\sigma & = & \int \sigma(E) \, dE 
= \frac{2J_{IMS}+1}{2J_0+1} \,
\left(\frac{\pi \hbar c}{E_{IMS}}\right)^2 \, \times \nonumber \\
 & & \, \, \times \, \frac{\Gamma^\ast_{IMS \rightarrow
    {\rm{low-}}K}\, \Gamma^\ast_{IMS \rightarrow {\rm{high-}}K}}{\Gamma^\ast}
\label{eq:isig}
\end{eqnarray}
$\Gamma^\ast_{IMS \rightarrow{\rm{low-}}K}$ and $\Gamma^\ast_{IMS
  \rightarrow{\rm{high-}}K}$ are the total decay widths from the IMS to
\loK\ and to \hiK\ states under stellar conditions (including all cascades),
$\Gamma^\ast = \Gamma^\ast_{IMS \rightarrow{\rm{low-}}K} + \Gamma^\ast_{IMS
  \rightarrow{\rm{high-}}K}$ is the total decay width, $J_{IMS}$ and $J_0$ are
the spins of the IMS and the initial state, and the energy $E_{IMS}$ is given
by the difference between the excitation energies of the IMS and the initial
state: $E_{IMS} = E_x(IMS) - E_0$. The factor $\Gamma^\ast_{IMS
  \rightarrow{\rm{low-}}K} \times \Gamma^\ast_{IMS \rightarrow{\rm{high-}}K} /
\Gamma^\ast$ in Eq.~(\ref{eq:isig}) may also be written as $b^\ast_{IMS
  \rightarrow{\rm{low-}}K} \times b^\ast_{IMS \rightarrow{\rm{high-}}K} \times
\Gamma^\ast$ where $b^\ast_{IMS \rightarrow{\rm{low-}}K}$ and $b^\ast_{IMS
  \rightarrow{\rm{high-}}K}$ are the total decay branchings of the IMS under
stellar conditions.

It is important to point out that the total decay widths (including all
cascades) to \loK\ and \hiK\ states enter into Eq.~(\ref{eq:isig}). This is a
consequence of the thermal population of excited states under stellar
conditions; for details, see \cite{Mohr07,Mohr06}.

The stellar reaction rate $\lambda^\ast$ in Eq.~(\ref{eq:lam}) is given by the
sum over the integrated cross sections $I^\ast_\sigma$ of all IMS where the
contribution of each IMS is weighted by the number of thermal photons at the
corresponding excitation energy. Because of the exponential dependence of the
thermal photon density in Eq.~(\ref{eq:planck}), practically only very few
low-lying IMS do contribute to the sum in Eq.~(\ref{eq:lam}). In the
present study we restrict ourselves to the experimentally confirmed IMS
in \luN\ at 839\,keV and a further candidate at 725\,keV \cite{Gin09}; for
\taN\ we analyze the lowest IMS candidate at 594\,keV \cite{Mohr07}.

The stellar reaction rate $\lambda^\ast(T)$ is strongly temperature dependent
because of the roughly exponential factor $E^2/[\exp{(E/kT)} - 1]$ in
Eq.~(\ref{eq:planck}). In addition to this explicit temperature dependence
there is further implicit temperature dependence of $\lambda^\ast(T)$ because
the widths $\Gamma^\ast$ in Eq.~(\ref{eq:isig}) also depend on
temperature. This further temperature dependence will be discussed in detail
in the next Sect.~\ref{sec:mod}; see also Eq.~(\ref{eq:gammatemp}).

For the sake of clarity we will use the symbol $\lambda^\ast$ in units of
s$^{-1}$ only for the stellar reaction rate between \loK\ and \hiK\ states in
Eq.~(\ref{eq:lam}); the symbol $\lambda$ will be used for transition rates 
between levels or groups of levels (in the same $K$ group). Levels will be
further characterized by their lifetimes $\tau$ instead of their decay
constants $\lambda = 1/\tau$. All energies are given in keV.

\subsection{Transitions in \luN\ and \taN }
\label{sec:trans}

\subsubsection{\luN}
\label{sec:trans176}
\begin{figure}[thbp] 
  \centering
  \includegraphics[width=7.4cm]{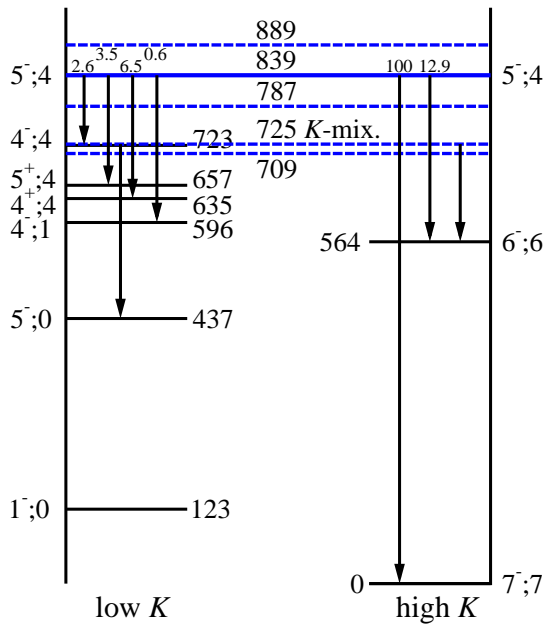}
  \caption{ 
   (Color online)
   Partial level scheme of \luN\ with \loK\ states on the left and
   \hiK\ states on the right. IMSs are indicated by blue lines over the full
   width of the diagram. The IMS at 839\,keV (full line) decays to \loK\ and
   to \hiK\ states. Relative $\gamma$-ray branches $b^{\gamma,{\rm{rel}}}$
   normalized to the dominating ground state branching
   $b^{\gamma,{\rm{rel}}}_{839 \rightarrow 0} = 100$ are given for the IMS at
   839\,keV.  \Kmix\ of two neighboring $7^-$ levels at 724.7\,keV and
   725.2\,keV may lead to a further IMS \cite{Gin09}. New low-lying IMS have
   been identified in a $K=4$ band at 709\,keV, 787\,keV, and 889\,keV
   \cite{Dra10}. The dashed lines indicate IMS that are not studied in detail
   in this work.  
   }
  \label{fig:lu176level}
\end{figure}

A simplified level scheme of \luN\ is shown in
Fig.~\ref{fig:lu176level}. There is an experimentally confirmed IMS at
839\,keV, and a further candidate for an IMS at 725\,keV has been suggested
from the almost degeneracy of a \loK\ $7^-$ level and a \hiK\ $7^-$ level
\cite{Gin09}. Very recently, new low-lying IMS have been found by coincidence
$\gamma$-spectroscopy \cite{Dra10}. 

Here we analyze the experimentally confirmed IMS at 839\,keV and its decays to
the \loK\ levels at 723\,keV, 657\,keV, 635\,keV, and 596\,keV and to the
\hiK\ levels at 564\,keV and 0\,keV (ground state). Further details of the
transitions are listed in Table \ref{tab:lu176trans}. There are
transitions in a wide range of energies for this IMS at 839\,keV. Thus,
conclusions can also be drawn for transitions from other IMS
\cite{Gin09,Dra10} without a further detailed analysis.
\begin{table}[htbp]
\caption{
Transitions in $^{176}$Lu (from \cite{ENSDF}).
\label{tab:lu176trans}
}
\begin{center}
\begin{tabular}{crcrcr}
\hline
$J^\pi_i;K$ & \multicolumn{1}{c}{$E_{x,i}$} 
  & $J^\pi_f;K$ & \multicolumn{1}{c}{$E_{x,f}$} 
  & transition & \multicolumn{1}{c}{$\Gamma^\gamma_{i \rightarrow f}$} \\
  & \multicolumn{1}{c}{(keV)} & & \multicolumn{1}{c}{(keV)} & &
\multicolumn{1}{c}{($\mu$eV)} \\ 
\hline
$5^-;4$ & 839 & $4^-;4$ & 723 & (M1) & 1.3\footnote{from $\Gamma^\gamma_{839 \rightarrow 0}$ and measured branching} \\
$5^-;4$ & 839 & $5^+;4$ & 657 & (E1)\footnote{tentative assignment} &
1.8\footnotemark[1] \\
$5^-;4$ & 839 & $4^+;4$ & 635 & (E1)\footnotemark[2] & 3.3\footnotemark[1] \\
$5^-;4$ & 839 & $4^-;1$ & 596 & (M1,E2)\footnotemark[2] & 0.3\footnotemark[1] \\
$5^-;4$ & 839 & $6^-;6$ & 564 & M1 & 6.5\footnotemark[1] \\
$5^-;4$ & 839 & $7^-;7$ & 0   & E2 & 50.0\footnote{assumed within the
  experimental errors; see text.} \\
$7^-;0$ & 725 & $5^-;0$ & 437 & E2 & 27.3\footnote{calculated value
  \cite{Gin09}}  \\
$7^-;6$ & 725 & $6^-;6$ & 564 & (M1) & 15.8\footnotemark[4] \\
\hline
\end{tabular}
\end{center}
\end{table}

A candidate for an IMS at 725\,keV has been suggested by \cite{Gin09}; the
suggestion is based on a theoretical study of \Kmix\ of two $7^-$ states at
724.7\,keV and 725.2\,keV with $K = 0$ and $K = 6$. The 725\,keV states decay
to the \loK\ state at 437\,keV and to the \hiK\ state at 564\,keV. 

Members of the $K=4$ band with its $4^+$ band head at 635\,keV have been
identified as IMS recently \cite{Dra10}. Weak branches to the \hiK\ $7^-;7$
ground state have been found for the $6^+$, $7^+$, and $8^+$ members of this
band at 709\,keV, 787\,keV, and 889\,keV. The main decay branch from this
band goes to the \loK\ side. From the estimated transition strengths in
\cite{Dra10} it results that only the lowest IMS at 709\,keV may have
significant influence on the stellar transition rate $\lambda^\ast$.

Unfortunately, the lifetimes of the two $7^-$ states at 725\,keV are unknown,
and only lower and upper limits for the lifetime of the $5^-$ state at
839\,keV are available in literature. For the following discussion we take
$\Gamma^\gamma_{839 \rightarrow 0} = 50$\,$\mu$eV that corresponds to a
partial lifetime of $\tau_{839 \rightarrow 0} = 13.2$\,ps. This value is in
the experimental limits $10\,{\rm{ps}} \le \tau \le 433\,{\rm{ps}}$ for the
lifetime of the 839\,keV state because this state predominantly (branching
$\gtrsim 80\,\%$) decays by the $839 \rightarrow 0$ transition. In agreement
with the theoretical arguments in \cite{Doll99} and the experimental
photoactivation yields \cite{Van00,Kn05} (see discussion in \cite{Mohr09}
where $\tau \approx 12$\,ps is suggested with an uncertainty of about a factor
of two) we use a value close to the upper experimental limit of the width (or
lower limit of the lifetime).

\subsubsection{\taN}
\label{sec:trans180}
\begin{figure}[thbp] 
  \centering
  \includegraphics[width=7.4cm]{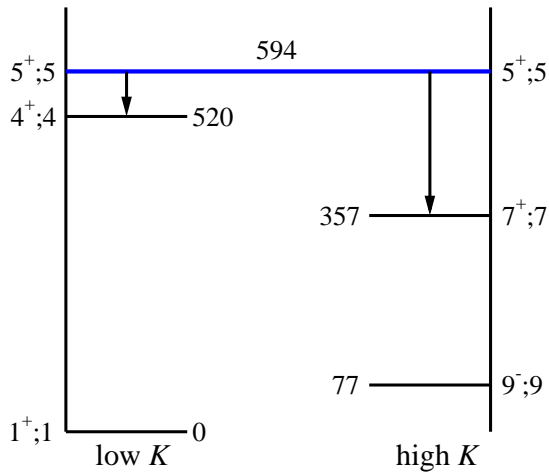}
  \caption{ 
   (Color online)
   Partial level scheme of \taN\ with \loK\ states on the left and \hiK\ states
   on the right. The IMS is indicated by a blue line over the full width of
   the diagram.
   }
   \label{fig:ta180level}
\end{figure}

Following \cite{Mohr07}, the lowest IMS in \taN\ is located at 594\,keV. It is
the band head of a $K = 5$ rotational band, and also the higher members of
this band have been assigned as IMS \cite{Wal01}. The 594\,keV level has a
half-life of $t_{1/2} = 16.1 \pm 1.9$\,ns and decays by a 72.2\,keV transition
\cite{Dra98,Sai99}, probably by a M1 transition to the 520\,keV level on the
\loK\ side. (Note that there is a surprising 2\,keV discrepancy in the
transition energy and the excitation energies that may be related to the
2\,keV shift of the $9^-$ isomer from $E_x = 75$\,keV in earlier compilations
to $E_x = 77$\,keV in the latest data base \cite{ENSDF}.) Based on reasonable
assumptions for the transition strength of the E2 transition from the 594\,keV
state to the $7^-$ state at 357\,keV on the \hiK\ side, it has been concluded
in \cite{Mohr07} that the 594\,keV state is the lowest IMS in \taN . A
simplified level scheme of \taN\ is shown in Fig.~\ref{fig:ta180level}.

\section{Modifications of transitions by the stellar plasma}
\label{sec:mod}
\subsection{Stellar transition rates and detailed balance theorem}
\label{sec:stran}
In this chapter, we have changed notations to have indices $I$, $L$, and $H$
to designate IMS, \loK , and \hiK\ states, respectively.  The stellar reaction
rate expression in Eqs.~(\ref{eq:lam}) to (\ref{eq:isig}) only includes
radiative excitation and spontaneous photon emission. In a stellar plasma at
thermodynamic equilibrium, induced photon emission has also to be included.
This can be easily done by changing Eq.~(\ref{eq:isig}) for a transition from
a \hiK\ state to a \loK\ state into:

\begin{eqnarray}
I^\ast_\sigma & = &  
 \frac{2J_{I}+1}{2J_H+1} \,
\left(\frac{\pi \hbar c}{E_{I}-E_{H}}\right)^2 \, \times \qquad
\nonumber \\
 & \times &
 \frac{ \Gamma^\ast_{IL} \Gamma^\ast_{IH}
   \frac{\exp{\left(\frac{E_{I}-E_{H}}{kT}\right)}}{\exp{\left(\frac{E_{I}-E_{H}}{kT}\right)}-1}}{\Gamma^\ast_{IL}
   \frac{\exp{\left(\frac{E_{I}-E_{L}}{kT}\right)}}{\exp{\left(\frac{E_{I}-E_{L}}{kT}\right)}-1}+{\Gamma^\ast_{IH}
     \frac{\exp{\left(\frac{E_{I}-E_{H}}{kT}\right)}}{\exp{\left(\frac{E_{I}-E_{H}}{kT}\right)}-1}}}
\label{eq:isigind}
\end{eqnarray}

However, it should be noted that $L$ and $H$ must designate single levels
here. When several \hiK\ levels or several \loK\ levels are involved, each
stellar transition rate must be dealt with separately.

Adding induced photon emission is only relevant when transition energies are
not too much larger than the plasma temperature $kT$. In the worst case that
will be presented below, a 72 keV transition in \taN\ at a temperature of
25 keV, the correction is only $5\%$. Thus, the approximation for the stellar
reaction rate in Eq.~(\ref{eq:lam}) remains valid for typical astrophysical
conditions. 

In a plasma at Local Thermodynamic Equilibrium (LTE), transition rates are
related to their corresponding inverse transition rates by the detailed
balance theorem. It can be easily proved that this still stands when dealing
with indirect (through the IMS) transition rates, so we can write:
\begin{eqnarray}
\frac{\lambda^\ast_{HL}}{\lambda^\ast_{LH}} &= &  \frac{2J_{L}+1}{2J_{H}+1}  \exp{\left(\frac{E_{L}-E_{H}}{kT}\right)} 
\label{eq:revers}
\end{eqnarray}
It is possible to define a global excitation and deexcitation rate when the
IMS state is excited from, or decays down to, a group of levels by summing over
the contributing levels $j$ \cite{Gos07}:
\begin{equation}
\lambda_{IL} = 
 \sum_j \lambda_{IL_{j}}
\label{eq:lbdil}
\end{equation}
and
\begin{equation}
\lambda_{HI} = 
 \frac{\displaystyle{\sum_j} \left(2J_{H_{j}}+1\right) e^{-\frac{E_{H_{j}}}{kT}} \lambda_{H_{j}I}}{\displaystyle{\sum_j} \left(2J_{H_{j}}+1\right) e^{-\frac{E_{H_{j}}}{kT}}}
\label{eq:lbdhi}
\end{equation}
%
These global rates do not verify the detailed balance theorem,
as no single energy and spin can be associated to the `global level'. The
detailed balance theorem can only be verified for a transition between two
individual levels, and not when some are grouped together into a global level.

However, in the case where one transition dominates all the other transitions 
from its group, the detailed balance theorem is approximately verified. In
particular, such is the case for \luN\ in this work.

\subsection{Modifications of transition rates by electronic environment}
\label{sec:modenv}
Electronic environment in stellar plasmas may influence decay or excitation
properties of nuclei. Internal conversion is strongly dependent on the number
of bound electrons, and nuclear transitions may be excited by its inverse
process NEEC \cite{Dzy07,Gos04}.

The huge number of low energy free electrons may also play a role in decay or
excitation by electron scattering \cite{Gos09} 
even though the transition rate is usually quite small
for high energy nuclear transitions. In the particular cases where an atomic
transition matches in energy a nuclear transition, NEET (Nuclear Excitation by
Electron Transition) and its reverse process BIC (Bound Internal Conversion)
become possible \cite{Mor04,Mor04b}. However, this last phenomenon is absent
for the nuclear transitions in \taN\ or \luN\ of this study as no
atomic transition matches the high energy nuclear transitions of interest.  

The net effect of all these processes is a modification of the excitation and
de-excitation rates leading to modifications of nuclear level lifetimes
\cite{Gos07}. All these processes have been dealt with under the LTE
hypothesis, which means that the detailed balance theorem can be used for each
individual process as well as for the total transition rate between two
levels.

The width $\Gamma^\ast_{i \rightarrow f}(T)$ for a transition from an initial
state $i$ to a final state $f$ under stellar conditions depends on temperature
and is given be the sum over several contributions:
\begin{eqnarray}
\Gamma^\ast_{i \rightarrow f}(T) & = &
\Gamma^{\gamma}_{i \rightarrow f} +
\Gamma^{IC}_{i \rightarrow f}(T) +
\Gamma^{(e',e)}_{i \rightarrow f}(T) \nonumber \\
& = &
\Gamma^{\gamma}_{i \rightarrow f} 
[ 1 + 
\alpha^{IC}_{i \rightarrow f}(T) +
\alpha^{(e',e)}_{i \rightarrow f}(T)
] \nonumber \\
\label{eq:gammatemp}
\end{eqnarray}
$\Gamma^{\gamma}_{i \rightarrow f}$ is the temperature-independent
$\gamma$-radiation width that is enhanced by the temperature-dependent widths
of conversion electrons $\Gamma^{IC}_{i \rightarrow f}(T)$ and of electron
scattering $\Gamma^{(e',e)}_{i \rightarrow f}(T)$. The $\alpha$ are the
corresponding dimensionless enhancement factors normalized to the radiation
width $\Gamma^{\gamma}_{i \rightarrow f}$. The $\alpha^{IC}_{i \rightarrow f}$
is the well-known internal conversion coefficient modified to take into
account the partial ionization of the atom and the modifications it induces on
the electronic wavefunctions.

The explanation of Eq.~(\ref{eq:gammatemp}) uses the standard wording for the
decay case. Although the underlying physics is exactly the same, the usual
wordings for the excitation case are ``nuclear excitation by electron
capture'' $\Gamma^{NEEC}$ instead of ``internal conversion'' $\Gamma^{IC}$ and
``inelastic electron scattering'' $\Gamma^{(e,e')}$ instead of ``superelastic
electron scattering'' $\Gamma^{(e',e)}$.

For completeness and clarification of the figures in Sect.~\ref{sec:res} it
must be pointed out that the radiation width $\Gamma^\gamma$ itself is
temperature-independent. However, the half-life (or decay rate) 
of a given state becomes
temperature-dependent at high temperatures because of induced photon emission
(see also Sect.~\ref{sec:stran}), even in the absence of the further
contributions of IC/NEEC and electron scattering  in
Eq.~(\ref{eq:gammatemp}).

\section{Results}
\label{sec:res}
As already mentioned in the introduction, plasma effects are important mainly
for transitions with low energies. Thus, capture reactions with typical
energies far above 1\,MeV are practically not affected in any astrophysical
scenario, whereas the production and destruction of isomers in the
astrophysical \spro\ has to be studied in detail. 

It is generally accepted that the astrophysical \spro\ operates in thermally
pulsing AGB stars \cite{Gal98,Buss99,Stra06}. In the so-called interpulse
phase neutrons are produced by the $^{13}$C\ran $^{16}$O reaction at
relatively low temperatures around $kT \approx 8$\,keV for about $10^4 -
10^5$\,years; this temperature is too low to affect isomer production and
destruction \cite{Mohr07,Mohr09}. During thermal pulses the $^{22}$Ne\ran
$^{25}$Mg neutron source is activated for a few years at temperatures around
25\,keV and densities of the order of $10^3$\,g/cm$^3$ \cite{Gal98}. For the
present analysis we adopt this density, and we study the temperature
dependence of various transitions in the chosen examples \luN\ and \taN .

The results are presented as temperature-dependent enhancement factors
${\cal{F}}(T)$ that relate the plasma effects (mainly NEEC and electron
scattering) to the effective radiative transition width
\begin{equation}
{\cal{F}}^X(T) = \frac{\Gamma^X(T)}{\Gamma^\gamma_{\rm{eff}}(T)}
\label{eq:enh}
\end{equation}
where the index $X$ stands for IC/NEEC, electron scattering, or NEET. The
presentation of the relative enhancement factor ${\cal{F}}$ instead of
$\Gamma^X(T)$ avoids complications for transitions with unknown radiation
widths $\Gamma^\gamma$. For $T \rightarrow 0$ the enhancement factors
${\cal{F}}$ in Eq.~(\ref{eq:enh}) are identical to the usual factors $\alpha$
in Eq.~(\ref{eq:gammatemp}).

It has to be kept in mind that the radiative width $\Gamma^\gamma$ in
Eq.~(\ref{eq:gammatemp}) is temperature-independent; but the radiative part is
enhanced by induced photon transitions at high temperatures leading to the
temperature-dependent effective radiation width $\Gamma^\gamma_{\rm{eff}}(T)$
in the denominator in Eq.~(\ref{eq:enh}):
\begin{equation}
\Gamma^\gamma_{\rm{eff}}(T)
= \Gamma^\gamma \left[ 1 + \frac{1}{\exp{(\Delta E/kT)}-1} \right]
\label{eq:gamgameff}
\end{equation}
The second part in the parenthesis is the enhancement due to induced photon
emission for a transition with energy $\Delta E$; see also
Eq.~(\ref{eq:isigind}) where the same factor was already used for the
definition of the integrated cross section $I_\sigma^\ast$. Obviously this
enhancement remains small at low temperatures and high transition energies,
i.e.\ $\Delta E \gg kT$.

All following results are presented within a range of temperatures from 1 keV to 
1 MeV. However, it should be noted that the results are non-relativistic estimates, 
which can lead to some errors for temperatures above a few hundred keV.

\subsection{Modification of widths in \luN\ and \taN }
\label{sec:resmod}
\subsubsection{\luN }
\label{sec:res176}
The lowest transition energy between the $5^-;4$ IMS state in \luN\ at
839\,keV and a lower state is 116\,keV. For such a high energy, one should not
expect the electrons to have a large influence on the transition rates.

\begin{figure}[htbp] 
  \centering
  \includegraphics[width=7.4cm,height=7.4cm,clip]{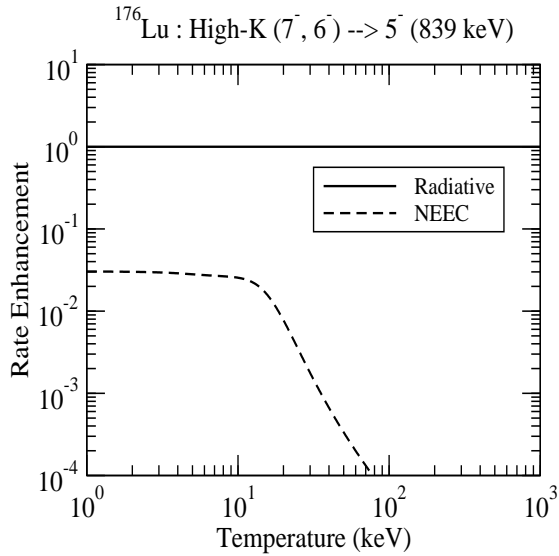}
  \caption{
    Transition rate enhancement factor for NEEC from \hiK\ levels to the IMS
    level of \luN\ at 1000\,g/cm$^3$. 
  }
  \label{fig:lu176rate1}
\end{figure}

First, we study the excitation of the $5^-;4$ IMS at 839\,keV from the
\hiK\ side, i.e.\ from the $7^-;7$ ground state and the $6^-;6$ state at
564\,keV. We plot the plasma enhancement factor as a function of
temperature for the chosen density of 1000\,g/cm$^3$ in
Fig.~\ref{fig:lu176rate1}. Only NEEC is not totally
negligible against radiative excitation, but it never amounts to more than a
few percents. 

\begin{figure}[htbp] 
  \centering
  \includegraphics[width=7.4cm,height=6.4cm,clip]{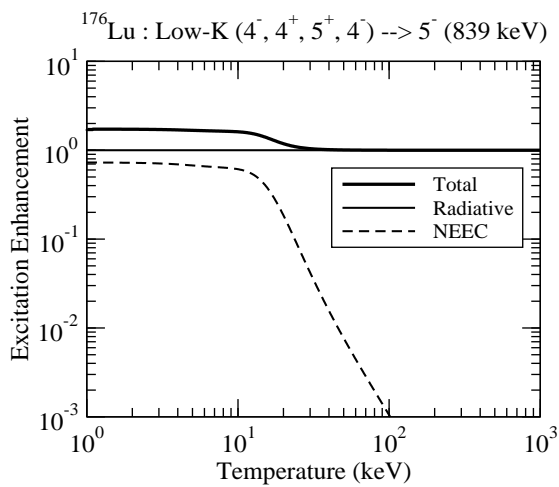}
  \caption{
    Transition rate enhancement factor for NEEC from \loK\ levels to the
    IMS level of \luN\ at 1000\,g/cm$^3$.
}
\label{fig:graf_ratio_lut176_nivL7}
\end{figure}

Excitations of the $5^-;4$ IMS at 839\,keV from the \loK\ side are somewhat
stronger affected. This is not surprising because of the lower transition
energies from the $4^-;1$, $4^+;4$, $5^+;4$, and $4^-;4$ states located between
596\,keV and 723\,keV. We find NEEC rates nearly equal to radiative rates for
temperatures lower than 10\,keV as shown in
Fig.~\ref{fig:graf_ratio_lut176_nivL7}. NEEC accounts for a global excitation
rate increase by a factor around 1.6 in this temperature range. 

\begin{figure}[htbp] 
  \centering
  \includegraphics[width=7.4cm,height=7.4cm,clip]{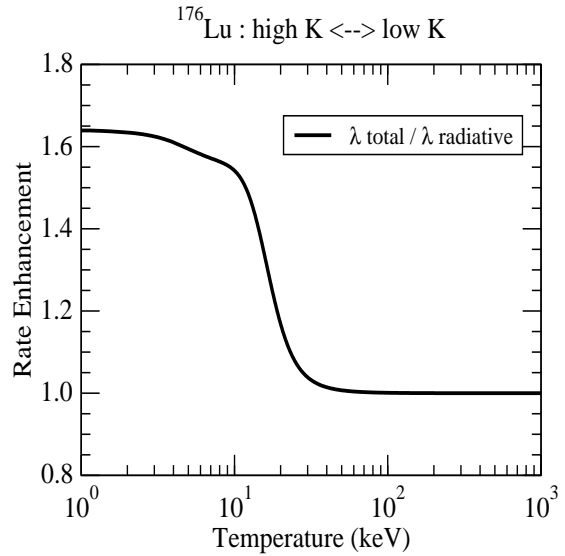}
  \caption{
    Stellar transition rate enhancement factor due to NEEC for \luN\ at
    1000\,g/cm$^3$. The enhancement at temperatures below
    15\,keV does not affect the nucleosynthesis of \luN\ in the \spro\ because
    the stellar rate drops below $10^{-15}$/s at 15\,keV.
  } 
  \label{fig:graf_ratioHL_lut176}
\end{figure}

This enhancement translates into the same factor on the stellar transition
rate Eq.~(\ref{eq:lam}) shown on Fig.~\ref{fig:graf_ratioHL_lut176}. However,
at temperatures below about 15\,keV the stellar transition rate from \hiK\ to
\loK\ states in \luN\ drops below $10^{-15}$/s or $3 \times 10^{-8}$ per year
\cite{Mohr09,Heil08},
i.e.\ it becomes negligible on the above mentioned timescale of a thermal
pulse. Consequently, the plasma modification of the stellar transition rate
does not affect the nucleosynthesis of \luN\ in the \spro .

\begin{figure}[htbp] 
  \centering
  \includegraphics[width=7.4cm,height=7.4cm,clip]{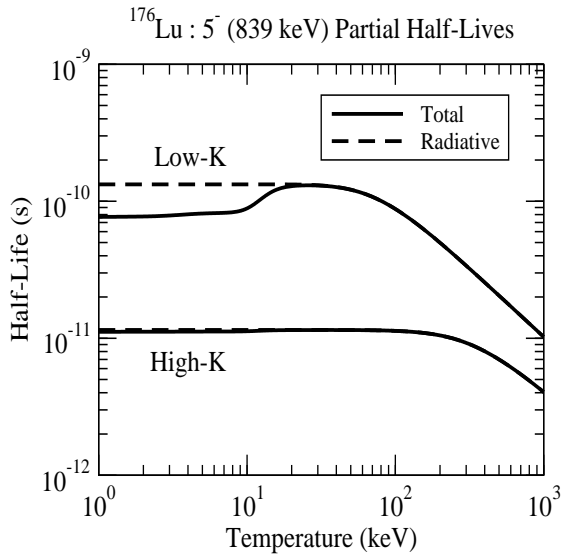}
  \caption{
    Partial half-lives of the $5^-;4$ IMS level of \luN\ towards \loK\ and
    \hiK\ levels at 1000\,g/cm$^3$. At low temperatures the branch to
    \loK\ states is enhanced by NEEC/IC. At high temperatures above 100\,keV
    induced photon emission shortens the half-lives.
  }
\label{fig:graf_tvie_lut176}
\end{figure}

The enhancement of the stellar transition rate is directly related to the
decrease of the partial half-life of the IMS level down to \loK\ levels as
shown on Fig.~\ref{fig:graf_tvie_lut176}. The dominating branch to the
\hiK\ side is practically not affected.

Two almost degenerate $7^-$ states around 725\,keV and their \Kmix\ have
been suggested as a further candidate for a low-lying IMS in
\luN\ \cite{Gin09}. The influence of the plasma environment on these two
almost degenerate $7^-$ states is small. The decay energies are 288\,keV for
the \loK\ branch and 161\,keV for the \hiK\ branch. These transition energies
are higher or at least similar to the transition energies in the \loK\ branch
of the $5^-;4$ IMS at 839\,keV that are enhanced only at very low temperatures
(see Fig.~\ref{fig:graf_ratio_lut176_nivL7} and discussion above). Thus, it
can be concluded that the IMS properties of the two $7^-$ states are not
affected by the plasma environment.

\subsubsection{\taN }
\label{sec:res180}

\begin{figure}[htbp] 
  \centering
  \includegraphics[width=7.4cm,height=7.4cm,clip]{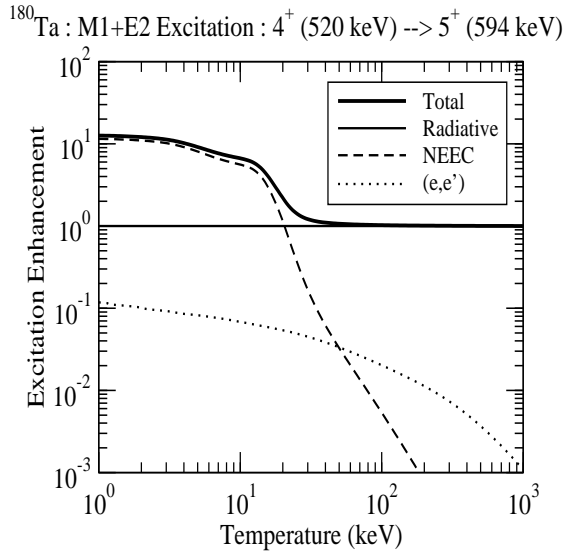}
  \caption{
    Transition rate enhancement factor for NEEC from the \loK\ $4^+$ level to
    the $5^+$ IMS level of \taN\ at 1000\,g/cm$^3$.
  }
  \label{fig:graf_ratio_tan180_niv12_decale}
\end{figure}

\begin{figure}[htbp] 
  \centering
  \includegraphics[width=7.4cm,height=7.4cm,clip]{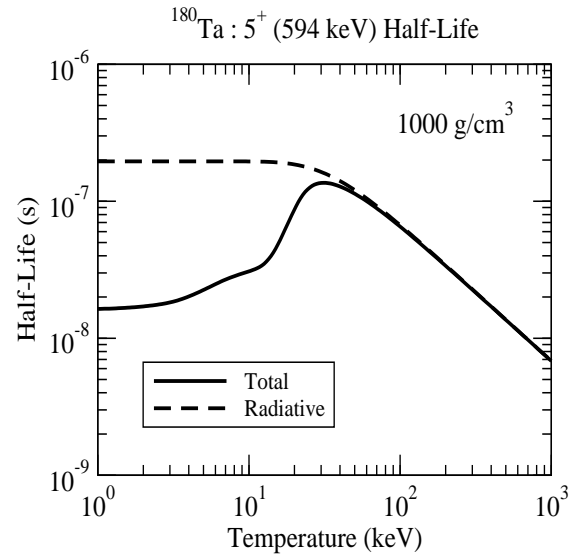}
  \caption{
    Partial half-life of the $5^+$ IMS level of \taN\ towards the $4^+$
    \loK\ level at 1000\,g/cm$^3$. The reduction of the half-life at low
    temperatures results from enhanced transitions by NEEC. The reduction at
    high temperatures is due to induced transitions.
  }
  \label{fig:graf_tvie_tan180}
\end{figure}

The candidate for the lowest IMS in \taN\ is a $5^+$ state at 594\,keV that
decays to the \loK\ branch by a 72\,keV (M1) transition; the laboratory
half-life is $t_{1/2} = 16.1 \pm 1.9$\,ns. Thus, at first glance, effects on
\taN\ appear to be stronger because of the relatively low transition energy of
only 72\,keV. Indeed, the excitation rate from the \loK\ 520 keV state
exhibits a large influence of electrons shown in
Fig.~\ref{fig:graf_ratio_tan180_niv12_decale}. For temperatures below 10\,keV,
electron inelastic scattering reaches $10\,\%$ of the radiative rate and NEEC is
10 times higher than the radiative rate. This factor can also be observed in
Fig.~\ref{fig:graf_tvie_tan180} with a factor of 10 decrease on the half-life
of the IMS level.

\begin{figure}[htbp] 
  \centering
  \includegraphics[width=7.4cm,height=7.4cm,clip]{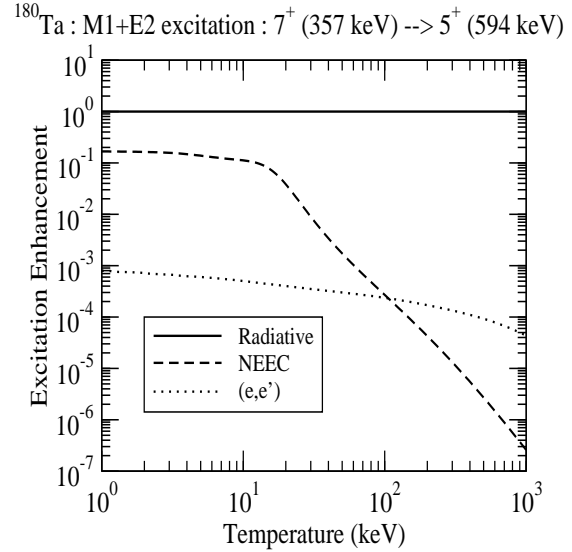}
  \caption{
    Transition rate enhancement factor for NEEC from the $7^+$ \hiK\ level to
    the $5^+$ IMS level of \taN\ at 1000\,g/cm$^3$.
  }
  \label{fig:graf_ratio_tan180_niv02_decale}
\end{figure}

The excitation rate enhancement for the 237\,keV E2 transition from the $5^+$
IMS to the \hiK\ $7^+$ state at 357\,keV is very small, even though in this
case NEEC is not the only contributor as electron inelastic scattering makes
an appearance as can be seen on Fig.~\ref{fig:graf_ratio_tan180_niv02_decale}.

\begin{figure}[htbp] 
  \centering
  \includegraphics[width=7.4cm,height=7.4cm,clip]{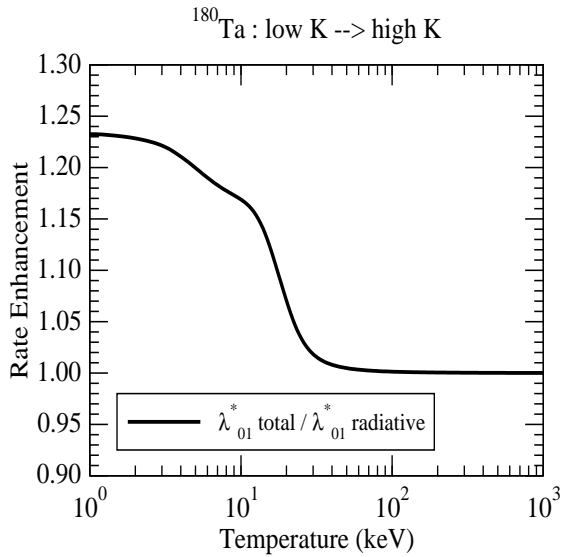}
  \caption{
    Stellar transition rate enhancement factor due to NEEC for \taN\ at
    1000\,g/cm$^3$.
  }  
  \label{fig:graf_ratio_tan180_niv01}
\end{figure}

Contrary to the \luN\ case, the rate enhancement of the \loK\ branch of the
IMS does not translate into a similar increase on the stellar transition rate
between \loK\ and \hiK\ states. Fig.~\ref{fig:graf_ratio_tan180_niv01} shows
that a $20\,\%$ increase can at best be expected for the lowest temperatures
because the excitation from the \hiK\ level is the relevant term in the
stellar transition rate. Similar to \luN , the small enhancement of the
stellar reaction rate at low temperatures below about 15\,keV does not affect
the nucleosynthesis in the \spro\ because the absolute rates are too small at
such low temperatures.

\subsection{Discussion of the results}
\label{sec:disc}

From the above shown examples it can be concluded that transitions with
energies above 200\,keV are practically not affected by the plasma environment
that is present under stellar \spro\ conditions. The influence of the stellar
plasma increases for lower transition energies and may reach about a factor of
two for transition energies above 100\,keV. Low-energy transitions below
100\,keV may change dramatically; e.g., a factor of about 10 has been found
for the 72\,keV transition in \taN .

NEEC is the main contributor to this increase with capture onto the $1s$ shell
amounting to the larger part. This effect disappears when the temperature
increases as free electrons have too much energy to be captured onto an atomic
shell.  The only other influence of electrons is inelastic
scattering. However, it is never greater than 10\,\% of the radiative
excitation rate or more than 1\,\% of the total transition rate. NEET remains
negligible as long as no matching transitions are present.

Changes in the strength of a particular transition do not directly translate
into modifications of the stellar reaction rate $\lambda^\ast$ for transitions
from the \loK\ to the \hiK\ levels. The stellar reaction rate $\lambda^\ast$
is proportional to the integrated cross section $I_\sigma^\ast$ in
Eq.~(\ref{eq:isig}) and thus proportional to a width factor 
\begin{equation}
\lambda^\ast \sim I_\sigma^\ast \sim \frac{\Gamma_1
  \Gamma_2}{\Gamma_1 + \Gamma_2}
\label{eq:gam1gam2}
\end{equation}
where the $\Gamma_i$ represent the \loK\ and \hiK\ branches under stellar
conditions. 
As long as one
of the partial widths dominates -- e.g.\ $\Gamma_1 \gg \Gamma_2$ and thus
$\Gamma = \Gamma_1 + \Gamma_2 \approx \Gamma_1$ -- this dominating width
$\Gamma_1$ cancels out in Eq.~(\ref{eq:gam1gam2}), and the stellar rate is
approximately proportional to the smaller width $\Gamma_2$. If the smaller
width corresponds to a $K$-forbidden transition with relatively high energies
above 200\,keV, then the stellar reaction rate $\lambda^\ast$
is practically not affected by
the plasma environment. This is the case for the decay of the lowest IMS in
\taN\ \cite{Mohr06} and also for the recently identified lowest IMS in
\luN\ \cite{Dra10}. 

Although \luN\ and \taN\ appear to have a very different behavior in terms of
modification of individual excitation rates by electrons, the global effects
on the stellar transition rates are very similar: a 20\,\% to 60\,\% increase
of the stellar rate is found for temperatures lower than 20\,keV. The major
change of the 72\,keV transition in \taN\ does not appear as a major
modification of the stellar reaction rate because this 72\,keV transition is
the dominating decay branch of the IMS in \taN .

\section{Summary and conclusions}
\label{sec:summ}
Under stellar conditions
the radiative transition width $\Gamma^\gamma$ for an individual transition
from an initial state $i$ to a final state $f$ is enhanced by electronic
transitions which are induced by the surrounding stellar plasma. The
enhancement factor ${\cal{F}} = \Gamma^\ast/\Gamma^\gamma_{\rm{eff}}$ is
composed of several effects. Under typical \spro\ conditions the dominating
effect is NEEC. Electron scattering plays a very minor role, and NEET remains
completely  negligible for practical purposes.

Typical \spro\ conditions are temperatures around $kT \approx 23$\,keV and
$\rho \approx 10^3$\,g/cm$^3$ for the helium shell flashes in thermally
pulsing AGB stars. Under these conditions we find negligible enhancement
factors ${\cal{F}} \approx 1$ for transitions with energies above $\Delta E =
150$\,keV. At energies around 100\,keV, ${\cal{F}}$ increases, but remains
below a factor of two. Further lowering of the transition energy down to about
50\,keV leads to dramatic enhancement factors up to one order of magnitude
(${\cal{F}} \approx 10$). Transitions with energies below 50\,keV are even
further enhanced; but nuclear transitions with such low transition energies are
very rare. 

The nucleosynthesis of \luN\ and \taN\ is affected by low-lying $K$-isomers in
these nuclei and the production and destruction of these isomers via
transitions to IMS. The stellar transition rates $\lambda^\ast$ for
transitions from \hiK\ to \loK\ states are defined by the decay properties of
the IMS, i.e.\ by a combination of the individual transition strengths. For
\luN\ the stellar plasma does not lead to a significant modification of the
stellar transition rate $\lambda^\ast$ because the lowest transition energy of
116\,keV is sufficiently high, and thus all individual transitions remain
unaffected by the plasma. For \taN\ a significant enhancement of more than a
factor of two is found for the low-energy $\Delta E = 72$\,keV transition from
the lowest IMS at 594\,keV. This low-energy transition is the dominating decay
branch of the IMS; but the stellar rate $\lambda^\ast$ is essentially defined
by the weak decay branch to the 357\,keV state (as suggested in \cite{Mohr07})
which remains unaffected because of its larger transition energy. Thus, more
or less by accident, the stellar rate $\lambda^\ast$ for \taN\ is not modified
significantly although one individual decay branch is modified by more than a
factor of two.

In summary, due to the plasma environment the stellar reaction rate
$\lambda^\ast$ for the production or destruction of $K$-isomers in \luN\ and
\taN\ does not change by more than about 20\,\% at \spro\ temperatures around
25\,keV and less than about 60\,\% at very low temperatures below
10\,keV. However, at these low temperatures the absolute rates are too low to
have influence on \spro\ nucleosynthesis; under these conditions,
corresponding to the long-lasting interpulse phase with $kT \approx 8$\,keV,
the \loK\ and \hiK\ states have to be treated as two separate species that are
practically decoupled because the IMS cannot be reached by thermal
excitations.

Within the present knowledge of IMS in \luN\ and \taN\ it may be concluded
that electronic effects due to the plasma environment do not play a relevant
role in the \spro\ nucleosynthesis of \luN\ and \taN . However, it should be
kept in mind that three new IMS (or a group of IMS) have been suggested in the
last few years: 
725\,keV \cite{Gin09} and 709\,keV, 787\,keV, and 889\,keV \cite{Dra10} in
\luN\ and 594\,keV in 
\taN\ \cite{Mohr07}. Each newly suggested IMS has its individual decay pattern
which has to be studied. It may have a weak low-energy branch that may be
significantly enhanced by the plasma environment. This low-energy branch may
finally define the stellar rate $\lambda^\ast$ according to
Eq.~(\ref{eq:gam1gam2}). So we conclude here that the plasma enhancement
should be taken into account for any low-energy transition below about
100\,keV.

\begin{acknowledgments}
We thank the participants of the ECT workshop {\it{International Workshop on
    Atomic Effects in Nuclear Excitation and Decay}} (ECT Trento 2009), in
particular Ph.\ Walker, G.\ D.\ Dracoulis, J.\ J.\ Carroll, F.\ G.\ Kondev,
for interesting and encouraging discussions, and the ECT for its kind
hospitality during the workshop.
\end{acknowledgments}

\end{document}